\documentclass[twoside,a4paper,11pt]{sea10}
\usepackage{graphicx}
\begin{document}
\pagenumbering{arabic}
\pagestyle{myheadings}
\thispagestyle{empty}
{\flushleft\includegraphics[width=\textwidth,bb=58 650 590 680]{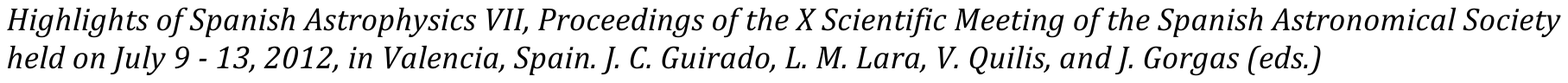}}
\vspace*{0.2cm}
\begin{flushleft}
{\bf {\LARGE
%
The Origin of Dwarf Early-Type Galaxies 
%
}\\
\vspace*{1cm}
%
Elisa Toloba$^{1,2,3}$
%
}\\
\vspace*{0.5cm}
%
$^{1}$
Departamento de Astrof\'{i}sica y CC. de la Atm\'{o}sfera, Universidad Complutense de Madrid,
              28040, Madrid, Spain\\
$^{2}$
UCO/Lick Observatory, University of California, Santa Cruz, 1156 High Street, Santa Cruz, CA 95064\\
$^{3}$
Observatories of the Carnegie Institution of Washington, 813 Santa Barbara Street, Pasadena, CA 91101
%
\end{flushleft}
%
\markboth{
The Origin of dEs
}{ 
%
E. Toloba  et al. 
%
}
\thispagestyle{empty}
\vspace*{0.4cm}
\begin{minipage}[l]{0.09\textwidth}
\ 
\end{minipage}
\begin{minipage}[r]{0.9\textwidth}
\vspace{1cm}
\section*{Abstract}{\small
%
We have conducted a spectrophotometric study of dwarf early-type galaxies (dEs) in the Virgo cluster and in regions of lower density. We have found that these galaxies show many properties in common with late-type galaxies but not with more massive early-types (E/S0). The properties of the dEs in Virgo show gradients within the cluster. dEs in the outer parts of the Virgo cluster are kinematically supported by rotation, while those in the center are supported by the random motions of their stars (i.e. pressure supported). The rotationally supported dEs have disky isophotes and faint underlying spiral/irregular substructures, they also show younger ages than those pressure supported, which have boxy isophotes and are smooth and regular, without any substructure. We compare the position of these dEs with massive early-type galaxies in the Faber-Jackson and Fundamental Plane relations, and we find that, although there is no difference between the position of rotationally and pressure supported dEs, both deviate from the relations of massive early-type galaxies in the direction of dwarf spheroidal systems (dSphs). We have used their offset with respect to the Fundamental Plane of E/S0 galaxies to estimate their dark matter fraction. All the properties studied in this work agree with a ram pressure stripping scenario, where late-type galaxies infall into the cluster, their interaction with the intergalactic medium blows away their gas and, as a result, they are quenched in a small amount of time. However, those dEs in the center of the cluster seem to have been fully transformed leaving no trace of their possible spiral origin, thus, if that is the case, they must have experienced a more violent mechanism in combination with ram pressure stripping, the open problem is that even galaxy harassment does not fully explain the observed properties for the pressure supported dEs in the center of the Virgo cluster.
%
\normalsize}
\end{minipage}
%
%
%
\section{Introduction}

Early-type galaxies (E/S0) are the most common type of object in areas of high density, while late-type galaxies dominate the population in the field \cite{Dress80}. This morphology segregation found in the distribution of galaxies in the Universe puts strong constraints in their formation and evolution. It points towards a major role of the environment in shaping galaxy evolution, but, the physical mechanisms involved in this process are not understood yet.

Dwarf galaxies are low mass and low luminosity systems that have low potential wells. If the environment plays an important role in the formation and evolution of galaxies, then, dwarfs would be the most affected systems. Within the dwarfs, dwarf early-type galaxies (dEs) are the most common population of galaxies in clusters \cite{FB94}, thus, they are excellent ''laboratories'' to test the mechanisms acting in these regions of high density.

Previous studies of dEs in clusters have opened a debate about the origin of these systems. Their seemingly simple and round appearance caused them to be classified as the low luminosity counterparts of massive elliptical galaxies, but there is not an agreement yet in the physical interpretation of the location of dEs in the scaling relations of E and S0 galaxies. While \cite{Kormendy09,Kormendy12} claim that E/S0 and dEs are two different populations of galaxies finding that they follow perpendicular trends in, e.g., the Kormendy relation, other works \cite{Graham03,Ferrarese06} interpret this same relation as a curvature in the properties of galaxies. The position of dEs on the Faber-Jackson and Fundamental Plane relations is also on dispute \cite{Geha03,DR05,Matkovic05,Kourkchi12a,Kourkchi12b}. These kinematic scaling relations 
have been traditionally used as a prove of galaxies following the same trends being the same family of objects, thus formed by the same processes \cite{Jorgensen96,P98}. More recently these diagrams have been related to the mass-to-light ratio of the galaxies \cite{Zaritsky06,Zaritsky08,Grav10,Zaritsky11}, thus, these relations also provide information about the galaxies' dark matter content.

Observational evidence has shown a wide variety of properties for dEs, something not common within the E/S0 population. Several photometric studies have found the presence of substructures in these galaxies, like disks, spiral arms, bars, lenses, irregular features \cite{Lisk06,Janz12}. Their surface brightness profiles follow nearly exponential laws ($S\grave{e}rsic$ $\sim 1$), instead of  de Vaucouleurs profiles ($S\grave{e}rsic$ $\sim 4$) as E galaxies \cite{Caon93,Ferrarese06,Lisk07}. Also, they are not composed of simple, old and metal rich stellar populations, but span a range in ages and are metal poor systems \cite{Mich08,Koleva09}. 

Not many mechanisms can explain together the morphology segregation and the heterogeneity of this family of galaxies. The purely internal processes, e.g. the kinetic energy generated in supernovae that sweeps away the gas of the galaxy leaving it without any fuel to create new stars \cite{YosAri87}, are not expected to correlate with the location of the galaxies in the Universe, thus, it can not explain why dEs are found mainly in clusters. Then, it seems more likely that external processes related to the perturbations induced on galaxies by the environment where they reside, are responsible for their formation.

There are two main mechanisms proposed for that transformation of properties: harassment, i.e. the gravitational interaction between close neighbours \cite{Moore98,Mast05}, and ram pressure stripping, i.e. the galaxy interaction with the intergalactic medium \cite{Boselli08a}. The predicted final properties of the transformed galaxies are different depending on the mechanisms affecting them. In a ram pressure stripping event the galaxy infalling the cluster lose its gas and rapidly quenches its star formation, but as this process does not directly affect the stars, their angular momentum should be conserved. Galaxy harassment is a significantly more violent process that can remove a large fraction of the stellar mass, change the morphology of the galaxy, and lose a significant fraction of the angular momentum of the stars. For a review on the topic see \cite{BG06}.

The kinematic analysis of dEs is therefore a powerful tool to identify the processes affecting galaxies in clusters. Several studies have conducted similar analysis finding that some dEs show significant rotation while others do not \cite{Ped02,Geha02,Geha03,VZ04,Chil09}. 

In this work we study a sample of dEs in the Virgo cluster, at different distances from its center, and in lower density environments with the aim of understanding the kinematic properties of these objects and distinguish between the scenarios proposed for their formation: ram pressure stripping and harassment.
We combine the kinematic information of these galaxies with photometry in the $K$ band, for the first time for dEs, and $V$ band to study the kinematic scaling relations: the Faber-Jackson and the Fundamental Plane relations. 

\section{Data}

\begin{figure}
\center
\includegraphics[scale=0.25]{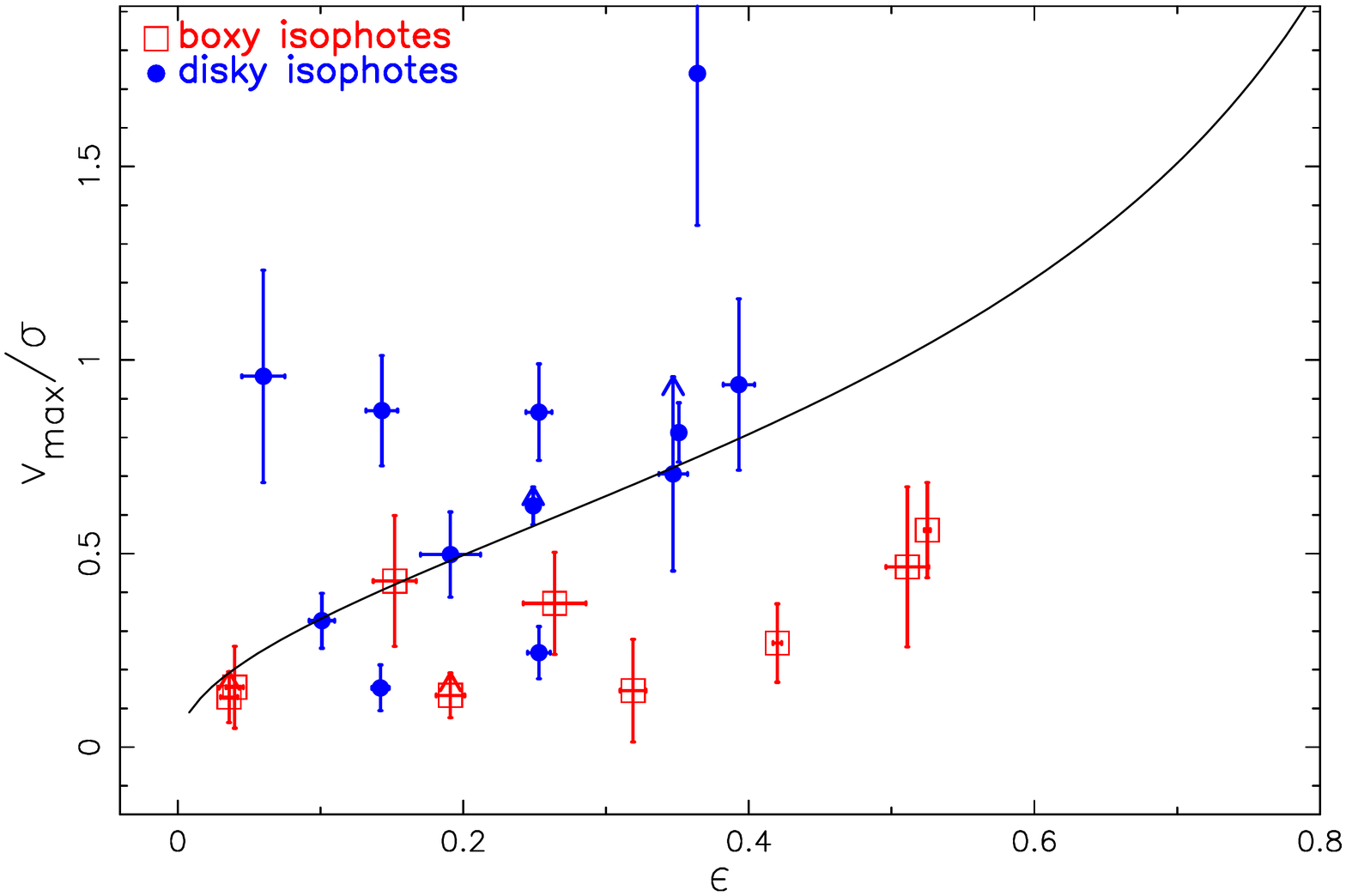} 
\includegraphics[scale=0.26]{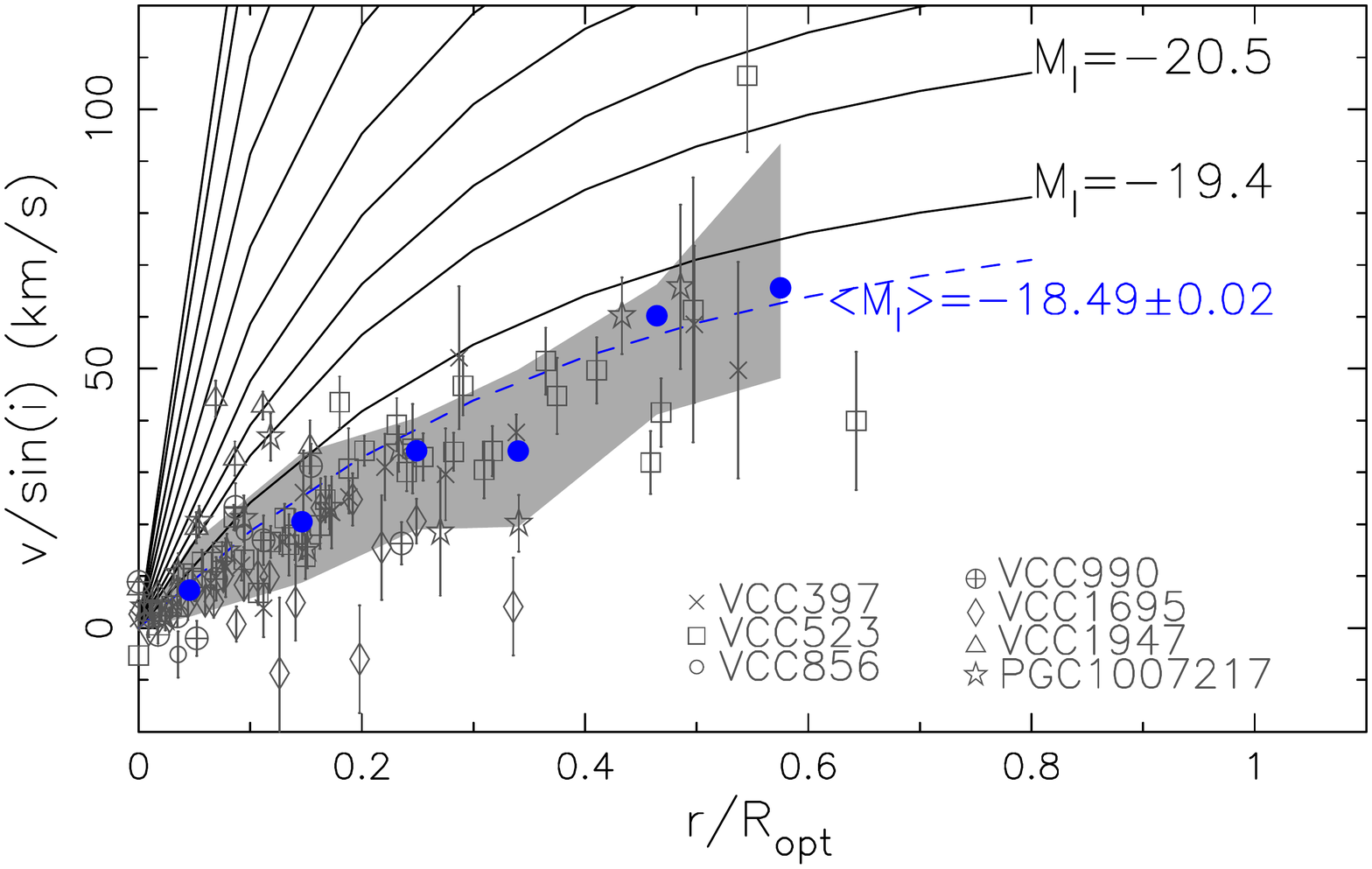} 
\caption{\label{F1} {\it Left Panel:} Anisotropy diagram from \cite{etj09b}. Blue dots and red squares indicate dEs with disky/boxy isophotes respectively. The solid line is the model for an isotropic oblate system flattened by rotation \cite{Binney78}. Lower limits on $v_{max}$ are indicated with arrows. {\it Right Panel:} Shapes of the rotation curves from \cite{etj11}. The observed rotation curves of rotationally supported dEs (galaxies over and above the solid line in the left panel, open symbols in this panel) are compared to the mean rotation curves of late-type spiral galaxies (black solid and blue dashed lines) from \cite{Cat06}. Blue filled dots represent the median observed rotation curve of rotationally supported dEs in bins of ${\rm r/R_{opt}}$, where ${\rm R_{opt}}$, the optical radius, is the radius that contains 83$\%$ of the light in the $I$ band. The grey area indicates rotation velocities within $1\sigma$ from the median.}
\end{figure}

The data used in this work is presented in \cite{etj09b,etj11,etj12}. It consists on a sample of 18 dEs in the Virgo cluster and 3 field dEs. The Virgo dEs were selected from the Virgo Cluster Catalog \cite[VCC]{Bing85}. They were chosen to have $ M_r>-16$, and dE or dS0 classification in the VCC catalog. The field sample of dEs was selected from SDSS \cite{SDSS} in the range $-18.5<rm M_r<-14.5$ and within a distance similar to Virgo, $5-25{\rm Mpc}$. Quiescent galaxies were selected assuming the color criterion ${\rm FUV-NUV}>0.9$ or $u-g>1.2$ when UV detections were not available.
All galaxies were selected to be within the GALEX MIS fields \cite{Boselli05}.

This work combines spectrophotometric information for each one of the 21 dEs that we observed in El Roque de los Muchachos Observatory (Spain) under the International Time Program ITP 2005-2007.
The spectroscopic data covers the wavelength range $3500-8950\AA$ with an intermediate resolution ($R \simeq 3800$). The details about the data acquisition, reduction and the kinematic measurements are described in \cite{etj11}. The photometric data were obtained in the $K$ band. 8 of the Virgo galaxies were not observed within our ITP program, therefore we used the $H$ band images from the GOLDMine database \cite{Gavazzi01,GOLDMine}. To undertake this analysis, we first had to homogenise the different sets of data. For those galaxies with only $H$ band information we assumed a color of $H-K=0.21$ \cite{Pel99}. To compare with less massive galaxies and to study possible trends due to stellar populations, we use the $V$ band. We have measured SDSS $g$ and $r$ photometry to convert them to $V$ using \cite{Blanton07}. The description of the observing strategy, the coherent calibration of all the images, and the measurement of the photometric parameters used throughout this work are detailed in \cite{etj12}. 

We use a Hubble constant of $H_0=73$ km s$^{-1}$ Mpc$^{-1}$ \cite{Mei07}, which corresponds to a distance for Virgo of 16.7 Mpc. The magnitudes are referred to Vega.

\section{Results}

The left panel of Figure \ref{F1} shows the anisotropy diagram. $v_{max}/\sigma$ is the ratio of the maximum rotation velocity measured in our rotation profiles and the velocity dispersion within the half light radius corrected for that rotation. Those galaxies with a radial coverage smaller than $6"$ appear with an arrow to indicate that the maximum rotation is likely not achieved. The colors indicate the shape of their isophotes in the near-infrared. We have compared this boxy/disky classification with the morphological classification performed by \cite{Lisk06} based on the unsharp masking technique. We find a good agreement between the two methods, which suggests that those galaxies with disky isophotes have underlying disky features such as spiral structures, disks or irregularities, while those dEs with boxy isophotes are smooth and regular dEs without any kind of substructure. The solid line in the Figure is the model for an isotropic oblate system flattened by rotation \cite{Binney78}. We will refer to rotationally supported dEs to those objects over or above the line, and pressure supported dEs to those below the line. Using the luminosity-weighted ages measured for these dEs by \cite{Mich08} and their projected distances with respect to M87, the center of the Virgo cluster, we find that the red squares are old galaxies (8 Gyr old on average) located in the inner $2^{\circ}$, while the blue dots are young dEs ($\sim 3$ Gyr younger than the red squares) located between  $2^{\circ}$ and $6^{\circ}$ from the center of the cluster.

\begin{figure}
\center
\includegraphics[scale=0.75]{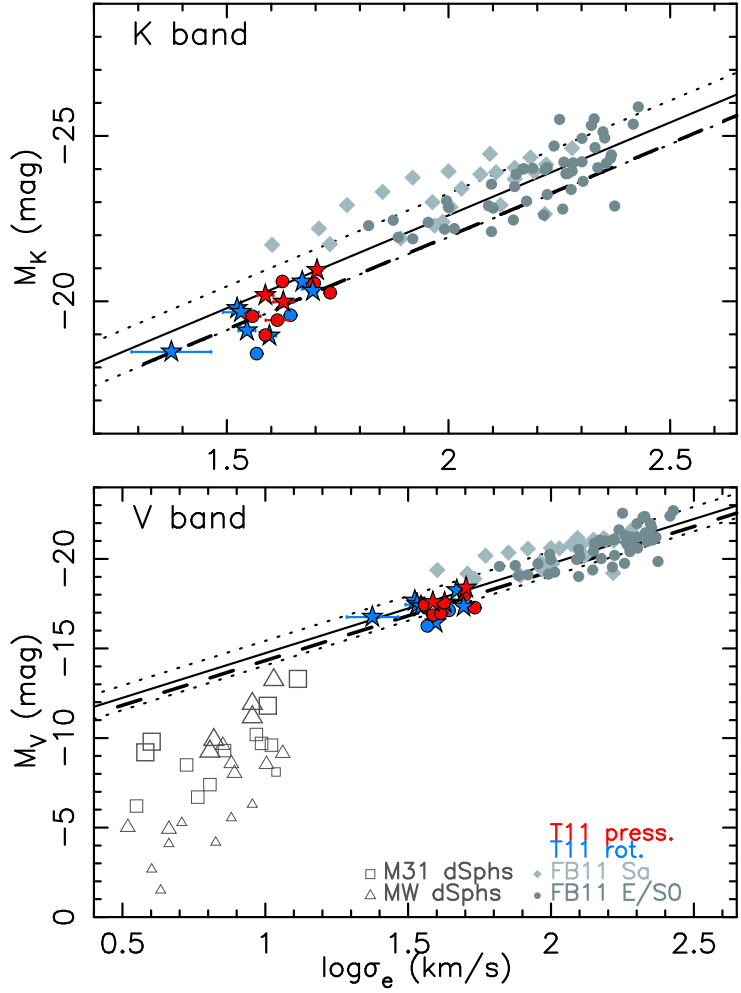} 
\caption{\label{F2} Faber-Jackson relation in the $K$ and $V$ bands from \cite{etj12}. Rotationally (in blue) and pressure (in red) supported Virgo dEs are compared to massive early-type and Sa galaxies from \cite{FB11}, Milky Way and M31 dSphs from \cite{Wolf10,Brasseur11,Tollerud12}. The size of the symbols for the dSphs are coded according to their dynamical mass-to-light ratio by \cite{Wolf10,Tollerud12} (the biggest symbols are for dSphs with ${\rm M_{dyn,1/2}/L_{1/2}}<50 {\rm M_{\odot}/L_{\odot,V}}$, the intermediate size are for those with $50 < {\rm M_{dyn,1/2}/L_{1/2}}<500 {\rm M_{\odot}/L_{\odot,V}}$, and the smallest for ${\rm M_{dyn,1/2}/L_{1/2}}>500 {\rm M_{\odot}/L_{\odot,V}}$). Solid and dotted lines are the fit and $1\sigma$ deviation for E/S0 by \cite{FB11}. The dashed line is the average perpendicular distance of T11 dEs to this fit.}
\end{figure}

The right panel of Figure \ref{F1} compares the shape of the rotation curve of the rotationally supported dEs in our sample with disk galaxies of different luminosities. The grey symbols are the different dEs from our sample, and the blue dots and grey area are the median rotation curve and $1\sigma$ deviation of all of them. The solid lines are the rotation curves for disk galaxies by \cite{Cat06}. The dashed blue line is the extrapolation of the parametrized rotation curves of disk galaxies to the average $I$ band magnitude of the rotationally supported dEs. It is remarkable the good agreement between the average rotation curve of rotationally supported dEs and the shape and amplitude of the rotation curve of disk galaxies of the same luminosity. These rotationally supported dEs also follow the same Tully-Fisher relation of massive late-type galaxies \cite{etj11}.

\begin{figure}
\center
\includegraphics[scale=0.5]{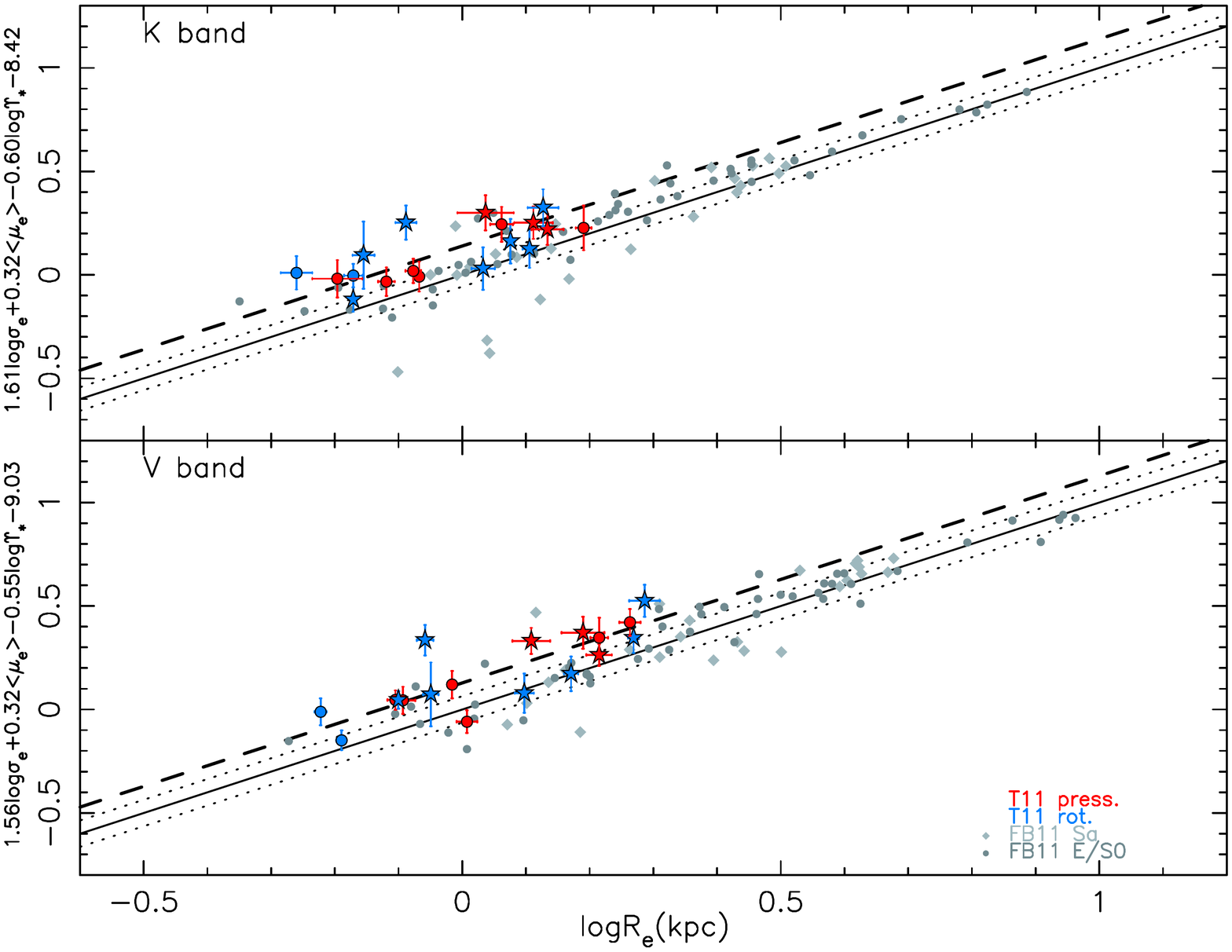} 
\caption{\label{F3} Fundamental Plane corrected for stellar mass-to-light ratio $(\Upsilon_*)$ from \cite{etj12}. Symbols and lines as in Figure \ref{F2}.}
\end{figure}

Figures \ref{F2} and \ref{F3} show the kinematic scaling relations, the Faber-Jackson relation (FJ) and the Fundamental Plane (FP). They are both in the $K$ and $V$ bands to compare the possible influence of the stellar populations on these relations, because the $K$ band measures the bulk of the stellar population. The filled dots and diamonds in these diagrams are the sample of massive early-type galaxies and Sa galaxies from \cite{FB11}. The solid line and dotted lines are the fit and $1\sigma$ deviation of E/S0 by \cite{FB11}. The red and blue symbols are our sample of pressure and rotationally supported Virgo dEs, respectively. The shape of their symbol indicates their luminosity-weighted age as determined by \cite{Mich08}. dEs older than 7 Gyr are indicated with dots, and dEs younger than 7 Gyr with asterisks. The black dashed line is the average perpendicular distance of our Virgo dEs to the fit for E/S0. In the lower panel of Figure \ref{F2} the open squares and triangles are the dwarf spheroidal galaxies (dSph) of M31 and the Milky Way \cite{Wolf10,Brasseur11,Tollerud12}.

Figures \ref{F2} and \ref{F3} show that the dEs have an offset with respect to the fit of E/S0 galaxies. This offset is independent from the band and also from the ages of the dEs and their kinematic support. In the FJ relation the deviation of the dEs is in the same direction of the dSphs. 

The mass-to-light ratio seems to drive the kinematic scaling relations \cite{Grav10}. When these relations are corrected for this ratio all objects, independently from their mass, follow the same fit \cite{Zaritsky06,Zaritsky08,Zaritsky11}. Therefore, if we correct for the stellar mass-to-light ratio and plot our dEs with respect to the fit of more massive early-type galaxies, then, the difference, if that exists, is a difference in the dark matter fraction because the total mass of a galaxy is the addition of two contributions, the stellar and dark matter.

As the Virgo dEs do not follow the same FP as the E/S0 in the same way they do not follow the FJ, we correct the FP from the stellar mass-to-light ratio. We make this correction in a similar way as it is done for the massive early-types. We assume a Kroupa initial mass function and generate a grid of models of exponentially declining star formation histories. The result is  $\Upsilon_*$ as a function of $H_{\beta}$. We apply this function to the $H_{\beta}$ values for our dEs by \cite{Mich08}. After this correction, there is still a clear offset with respect to the E/S0 galaxies. In \cite{etj12} we use this offset to estimate the dark matter fraction of these dEs with respect to massive early-type galaxies, and we find that dEs have a dark matter fraction of $\sim 42\%$. This value is in agreement with our previous virial mass estimation based on \cite{Cap06} in combination with simple stellar population modelling using the age and metallicity as derived by \cite{Mich08}. 

\section{Discussion and conclusions}

In this work we have addressed the question of where do dEs come from?. The debate between dEs being the low luminosity extension of more  massive elliptical galaxies \cite{Graham03,Ferrarese06}, and dEs being the evolved phase of late-type galaxies in a high density environment \cite{Kormendy85,VZ04,Kormendy09,Kormendy12} is still open.

Our kinematic analysis of a sample of dwarf early-type galaxies suggests that dEs have different properties from massive E/S0 galaxies. Those dEs that are rotationally supported rotate significantly faster than E/S0 galaxies, i.e. $v_{max}/\sigma$ is larger than the values found for E/S0 \cite{etj09b}. In \cite{Em07}, based on the analysis of 48 early-type galaxies, they found that none of them reaches $v_{max}/\sigma=1$, while half our sample does. In \cite{Cap12}, they analyse 260 early-type galaxies and find that there is a kinematic sequence with mass, the most massive early-type galaxies do not rotate while the less massive ones are the fastest rotators. However, there is a significant fraction of dEs, around half of our sample, that rotate very slow or do not show any signs of rotation, thus, they do not follow this same sequence with the stellar mass of the object.
Moreover, the scaling relations have been historically used as an observational constraint on the formation and evolution of galaxies. The tightness of these relations has been classically interpreted as an indication that early-type galaxies constitute a homogeneous population shaped by the same formation processes \cite{Jorgensen96,P98}. The fact that we find that dEs deviate from the Faber-Jackson and Fundamental Plane relations of more massive elliptical galaxies indicates that  their formation must have been different. In addition, we find that this deviation is in the same direction as dSphs, and for dSphs, the most dark matter dominated systems known, the distance to the scaling relations is related to their dark matter fraction \cite{etj12}. We have used their distance to the Fundamental Plane to measure a $\sim 42\%$ dark matter fraction of dEs in comparison with Es within their half-light radius.

On the other hand, there are many similarities between dEs and late-type galaxies. From a purely photometric point of view, both dEs and late-type galaxies have exponentially declining surface brightness profiles \cite{Caon93,Ferrarese06,Lisk07}. In addition, dEs are also a very heterogeneous group of galaxies, showing underlying structures like disks, spiral arms and irregular features, although there is a significant fraction of them that have smooth and regular surface brightness profiles \cite{Lisk06}. In the case of their stellar populations, dEs are not simple, old and metal-rich galaxies as E/S0, they expand a wide range in luminosity-weighted ages, from 1 to 12 Gyr, and they are metal-poor \cite{Mich08,Koleva09}. 
In our kinematic study of dEs in the Virgo cluster we have found that the rotational support of these galaxies depends on their distance to M87, considering this galaxy as the center of the cluster. Those dEs in the outer parts of the cluster are rotationally supported while those concentrated around M87 are pressure dominated. We have also found that the rotationally supported dEs have rotation curves with the same shape and amplitude of late-type spirals of the same luminosity, they have disky isophotes, disky underlying features, and their ages are around 3 Gyr younger than those dEs in the center of the cluster, which have boxy isophotes and do not have any underlying substructure.

From this study the differences between dEs and E/S0 galaxies and the similarities between dEs and late-type galaxies are highlighted. Therefore, it has helped us to have a better understanding of the origin of these systems. The correlation of the spectrophotometric properties of these dEs with their distance to the center of the Virgo cluster rules out an internal mechanism as the procedure to form this kind of galaxies (i.e. supernovae explosions sweep away the gas of the galaxy leaving it without fuel to create new stars \cite{YosAri87}), and it points towards an environmental process. Within the different external processes, ram pressure stripping explains all the properties of the rotationally supported dEs. In a ram pressure stripping event, the late-type galaxies infalling into the cluster lose their gas and become quiescent galaxies quite rapidly just through their interaction with the intergalactic medium \cite{Boselli08a}. This mechanism only affects, in the short term, the gas of the galaxy leaving the kinematic of the stars untouched, thus, one would expect that the kinematic properties of quiescent galaxies affected by ram pressure stripping to be the same as late-type galaxies of the same luminosity, which is what we have found for those dEs in the outer parts of the Virgo cluster. However, those dEs surrounding M87 are more difficult to explain. Their lack of rotation and substructures suggest that they might have suffered a more violent event than just ram pressure stripping, but, their older ages also suggest that they have been longer in the cluster, thus their several passes through the center could have shaped them, but, which is the physical mechanism of a transformation in a very high density region?. Harassment, the interaction of a galaxy with its neighboring objects, is the main candidate. It is significantly more violent than ram pressure, but simulations of galaxy harassment suggest that this process is only effective near the cluster center. \cite{Smith10} simulates the outer parts of the Virgo cluster and finds that harassment is not strong enough to transform low luminosity late-type galaxies into dEs because of the lack of a high number of neighbours. In the center of the cluster this mechanism is much more efficient but still it is not enough as to explain the properties of dEs nearby M87. In the simulations of \cite{Moore99,Mast05} although $\sim 90\%$ of the stars are tidally stripped in the interaction, a strong underlying spiral structure remains after the event, while no substructure if found in the dEs located in the central parts of the cluster. Thus, it looks like a combination of ram pressure stripping and harassment is helping to shape these non-rotating dEs in the center of the cluster, but a piece of the puzzle is still missing to fully explain the spectrophotometric properties of dEs in the center of the Virgo cluster.

In summary, many are the properties that make dEs look alike to late-type galaxies. The kinematic properties of dEs in the Virgo cluster suggest that these are late-type galaxies that entered the cluster and have been transformed into quiescent objects. The physical mechanisms that make the transformation are clear for those dEs found in the outer parts of the Virgo cluster, their properties agree with a ram pressure stripping event. However, the characteristics found for dEs in the center of the cluster make their origin more intriguing. A combination of ram pressure and harassment has affected these galaxies in their travel through the cluster, but, these mechanisms are not enough to explain their lack of rotation and underlying substructure. We are conducting new observations with a larger coverage in radial kinematic profiles to throw some light in this direction.

\section*{Acknowledgments}   
ET acknowledges the SEA committee for awarding this work with the prize for the best PhD Spanish dissertation  defended on 2011. ET is very thankful for having been invited to present her work at the 2012 Spanish Astronomical Society meeting celebrated in Valencia. ET is in debt to her PhD supervisors, A. Boselli and J. Gorgas, and also to R.F. Peletier for all the very useful discussions, suggestions and their hard work on the development of this project.
%

%
\end{document}